# A decision problem for bank branch site selection: A GIS Mapping perspective with Maximal Covering Location Problem: A case study of Isfahan, Iran.

Zahra Namazian[a*], Emad Roghanian[b]

[a*]MSc. Graduate, Department of Industrial Engineering, K. N. Toosi University of Technology, email: znamazian@mail.kntu.ac.ir.

[b] Associate Professor, Department of Industrial Engineering, K. N. Toosi University of Technology, email: e_roghaniaan@yahoo.com.

**Abstract**

One of the most significant decision-making processes is the site selection. If the site correctly selected, the access for the best customers as well as the greatest market potential would be guaranteed. Bank branch optimal location is one of the most significant strategic issues in the competitive market especially for private banks because of the global competition and high customer expectations. This study presents a Geographic Information System (GIS) based model for locating suitable sites for making new branches by using data sources. Also, Maximal Covering Location Problem (MCLP) was used to select branches that the maximum demand might be reached within a pre-specified target travel time. The model was implemented for Ansar bank in Isfahan, Iran. The criteria were restricted through demographic attributes, competition, transportation, flexibility and cost in GIS mapping approach. Finally, the results illustrated the efficiency and applicability of the proposed integrated method.

## 1. Introduction

Selecting an appropriate location is considered as an important decision since poor site selection would be costly and hard to reverse (Ernst et al., 2018). Inappropriate choice of a site location would lead to higher transportation costs and loss of qualified labor. The aim for finding an appropriate site is to select the best location that meet the predetermined selection criteria (Zanjirani Farahani, 2012). There are various factors significantly affecting our daily life and businesses; however, several papers have proposed models and studied the effectiveness of these factors (Rezazadeh et al., 2018). The site selection problem is not new among the research community; the problem has inspired a rich and ever-growing body of literature (Liu and Hodgson, 2013).

The site selection problem prevails in the fiercely competitive business environment. U.S. companies spent around 8.7% of the U.S. Gross Domestic Product (GDP) (i.e. $910 billion) for business logistics systems in 2002 (Shi et al, 2004). The growth over the past decade came from the non-Jordanians, who represent around one third of the Kingdom's population now as Department of Statistics (DoS) estimated their number at 2.9 million. Nearly half, or 49.7%, of non-Jordanians in the Kingdom live in Amman, which is also home to 38.6% of Jordanians. The growth would definitely affect the population of the country as well as building densities (Potter et al., 2007).

Nowadays, service organizations have significant effects on the countries' economic development. Regarding the importance of financial activities, banking industry as a service industry, holds a great potential in business development. As the human body cannot live without blood, the economy also cannot flourish without banks. Banks help to maintain the flow of cash in the economy by implementing the policies. Therefore, banks play a significant role for the growth of economy. Also, banking industry is progressively developing to move toward the path of privatization and to provide a competitive environment. Thus, to sustain in the uncertain marketplace, the banks should meet and satisfy their customers' requirements (Tavakoli and Shirouyehzad, 2013). Also, convenience is one of the most important factors for selecting the branch of banks with the customers in a specific area (Adil, et al, 2013).

The maximal covering location problem (MCLP) deals with the problem of selecting an optimal location of a number of existed facilities within a set of customers. Since each customer has a specific demand, the facilities have to be located to cover demands of customers by the facilities (Atta et al, 2018). Covering problems constitute the important family of facility location problems with widespread applications. These problems embed a notion of coverage radius or proximity (distance or travel time between points) that shows whether a given demand point can be served or "covered" by a potential facility location (Madani et al, 2017). A demand point is then said to be covered by a facility if it lies within the coverage radius of this facility (Cordeau et al, 2019). To cover demand point, this model finds the maximum covered business potentials subject to some constraints. Therefore, in early years, there are several studies for extensive reviews to demonstrate various types of problems related to maximum coverage location problem (MCLP) since it is one of the famous locating public service facilities to cover the demand (Bansal, 2018).

GIS is a vital location analytic method which is able to support map-based display along with geographic data creation, manipulation, and management. It helps to provide functions for suitability evaluation. Site selection illustrates measuring the needs of a proposed facility against the merits of potential locations (Vahidnia et al., 2008). To understand the effects of the various important factors, which have profound effects on the locating places, the GIS mapping is very useful. Rajendran and Gambatese (2009) proposed that the best way to improve the sustainability of buildings is to consider the site safety from inception. This method would be useful to maximize the benefits of project objectives in terms of quality, cost, time, and safety (Rajendran, 2006). On the other hand, the analytical and practical questions can be answered by using different GIS tools and functions. The simplest GIS tools that can be used at every GIS application are related to querying spatial, displaying, and attribute data. Therefore, most studies use GIS software which should have at least six main modules that are known as data input, data model, data storage, data manipulation, data analysis, and data output module. For example, ArcGIS software, which is one of the most significant and useful software, is designed to have these modules with various desktop applications known as ArcCataloge, ArcMap, ArcGlobe, ArcToolbox and ModelBuilder. Using these applications, users can perform any GIS task, from simple to advanced, including mapping, geographical analysis, data editing and completion, data management, visualization and geoprocessing (Murad, 2011). It can illustrate attributes in relation to lines, points, and polygons, or otherwise thematic mapping (Grimshaw, 2000). Thus, through the literature, there are several studies consider their widespread applicability ranging public service facilities through GIS to solve covering problems which are discussed in data mining, machine learning, and bioinformatics (Bansal and Kianfar, 2017, Bansal, 2018).

According to above mentioned explanations, recent political events in Iran have made Isfahan an appealing destination for tourists, nation and investors due to its central location in the country, economic stability, and high security level. A report published by the Department of Statistics (DoS) on the Isfahan municipality, carried out by the department in 2014. For example, for a retail chain to open a new outlet, a manufacturer to choose the position of a warehouse, and a banker to select the locations for service points, it is necessary to make good decisions for the location and allocation of facilities. Hence, several studies have been considered for the location selection of banks based on different methods such as Geographic Information Systems (GIS), Maximal Coverage Location problem (MCLP) and Decision Support Systems (DSS).

The current paper targets the problem of banking sector. The problem is finding the best site for location of bank branches. It is a significant concern for the banks to find the best and optimum location for branching to increase their profitability and market share. Covering consumer demand is the other important factor that bank branches are faces with. The aim of this paper is to discuss a GIS application created for locating the branches of bank in Isfahan, Iran. In order to build this application, several data sets were collected including detailed information about each bank branches were covered in this application. Arc GIS software has been used to build a geo-database that covers all of the collected data for this application. Therefore, the paper maximizes the covering locations and solves the MCLP by analyzing GIS maps to find factors. It can be said that GIS is applied at several location studies to show the factors visually. In addition, it weights these factors by using ANP approach. This paper would discuss how to use GIS for branches of bank location assessment in Isfahan, Iran. After reviewing the literature, the GIS maps demonstrate various factors and layers. The fourth section presents maximum covering location problem. Finally, implications, limitations and conclusion would be discussed.

## 2. Literature Review

Recently, some studies in the banking community have focused on the potential applications of location-allocation problems and models for finding the best bank-branch location and optimize the allocation of demand to those locations. There are several studies which discuss Solution techniques for location-allocation models, and have been formulated these problems in continuous and discrete space. Therefore, the bank branch location studied by Fung (2001) indicated that costumers choose their banks based on their convenient way to reach the bank. Since estimating the feasibility of a bank branches development project and plan, defining the availability of the demand area is a first challenge (Garegnani et al., 2018). Dick (2008) suggested the local branch density and depositors' value geographic reach. Location selection is considered as a significant matter for companies since it would be a costly mistake that results in unnecessary relocating expenses, endeavor and competitive advantages losses, especially when the decision problem depends on the locating facilities (Palomino Cuya et al., 2013). On the other hand, the problem of deciding on the best site for a new branch may be viewed as part of the general problem of restructuring the bank branch network (García Cabello, 20017). Thus, Finding the best site for branches of a new may be viewed as part of the general problem of restructuring the bank branch network since it may influence by a need arises in the event of changes in bank regulations, motivating mergers that necessitate a redesign of the branch network (Kamble, 2011). Banks, credit and financial institutions to attract more financial resources need to select suitable locations for construction of their branches. Hence, the optimal location of banking facilities is a critical decision since the investment of a banking facility is usually over millions of dollars. A wrong decision of branch location may induce serious investment losses.

Geographical information systems (GIS) are used at several planning applications including health care planning, land use planning, and transportation planning. Locating is considered as one of the planning fields that could benefit from using this service technology. So, the use of GIS in modeling demand flows to the location of important facility. It is also used to show the interaction between location and the factors related to the specific area. The results are influenced by the attractiveness of retail location, the size of demand, and the cost of travel between location j and area i (Murad, 2011). Therefore, GIS technology is appropriate for a variety of usages including business planning, land surveying, and resource management. For example, businesses use GIS technology to delivering better services, solve problems, making good decisions, and find solutions for marketing,

In addition, GIS technology has been implemented in service management for displaying large volumes of diverse data pertinent to various local and regional planning activities. In addition, finding the best location faced considerable number of problems including sales, constructing demographic, and competitive analysis. It also faced to create effective marketing campaigns, schedule and route deliveries, and provide better customer care information system (Murad, 2011). Thus, extensive studies have been reported associated with GIS based facility locations and covering location problems (Wang, eta al, 2016). Vlachopoulou etal. (2001) used geographic information system to warehouse site selection decision. Cheng eta al, (2007) investigated the problem of determining optimum number and locations of ATM's by the help of GIS and covering models. Lyu et al (2018) demonstrated flood risk assessment in metro systems of mega-cities using a GIS-based modeling approach. Flood disasters have swept billions of dollars of property damages and deaths along with high waters, causing inundation of underground infrastructures (e.g., metro tunnels and facilities). Atta et al (2018) used genetic algorithm (GA)-based approach to utilize a local refinement strategy for faster convergence to solve MCLP. The genetic algorithm was applied on different MCLP instances, and it was proposed that the genetic algorithm with local refinement leads to better results in terms of percentage of coverage and time estimation to find the solutions in almost all the cases. Zabihi et al (2019) presented that by modeling and using GIS map, it is possible to estimate soil erosion, runoff and sediment in order to maintain and control the measures. Today, the use of (GIS) as a tool for modeling is common. GIS can be used to analyze geospatial factors and preserve process. Murray et al (2019) reviewed a class of location models and provided an overview of the methods used to solve these models. Therefore, GIS is an important method because of the broad use and application of location models to address the important problems and issues. To improve spatial analytical insights, the implications of models and methods are available through GIS. Case studies are offered to highlight ease of access to location models in GIS along with observed computational performance. Meyer (2011) examined the current location fire stations that deploy fire protection and emergency medical services in Toledo, Ohio. it was tried to improve the efficiency of coverage to decrease total travel times. The methodologies of the MINISUM location allocation strategy were employed and maximum distance restriction was used to exclude long and unacceptable response times, reduce response times, increase efficiency of emergency services, and consequently increasing effectiveness in service delivery. Alsalloum and Rand (2006) used Maximum Covering Location Problem to identify locations of emergency medical service stations that the maximum demand might be reached within a pre-specified target travel time A comprehensive example of the route/site selection process of a metro-rail network project was also presented. Chaudhary, et al, (2019) presented the importance of GIS in soil waste management by minimizing the cost, and maximizing the waste collection and transportation efficiencies in any areas. Ibrahim (2011) investigated the selecting of the best location for wastewater lift station of an under-construction industrial sewage system in El-Mahalla El-Kubra, Egypt. They used experts' judgments and fuzzy Analytical Hierarchy for weighting the criteria. Also, GIS was used to overlay and generate criteria maps and suitability map. The study ended with an assessment of proposed sites to the generated suitability map.

Site selection of service facilities has been considered as a vital concern in urban and regional areas, as well as in research topics for various disciplines such as geography, public policy, urban planning, engineering and operations researches (Yao, et al, 2019). There are several different studies that focused on the locating problems and proposed different approaches. Chauhan, et al (2019) presented that given a set of demand points and potential facility locations to cover the demand points and a set of fully available charged drones, an agency allocating a considerable number of capacitated facilities and assign drones to the located facilities to serve the demands.

The coverage objective provides more tools to evaluate the feasibility of using drones to deliver medical supplies such as defibrillators (Boutilier et al., 2017; Claesson et al., 2017), blood deliveries (Amukele et al., 2017) or critical relief after extreme natural events while accounting for drone battery range limitations. Also, gradual cover models considered the possibility of partial cover (Berman, et al, 2019). There are several studies on covering models, but the location set covering problem (LSCP) is the simplest one, and maximal covering model is the most popular one which seek to locate the minimum number of facilities that are necessary to cover all demand areas (Lotfalipour et al., 2014).

Government officials in Iran do not mostly tackle with using mathematical models for site location studies. Thus, the decision to build a bank is driven by using methods of GIS and MCLP. Hence, the paper was unable to get some of the required data to use implement more advanced choice models for this paper, and the research was limited to analyze and examining the current status of the bank locating. This is an opportunity for the officials in Isfahan to analyze location by GIS and ask for scientific studies as formal requirements to apply mathematical models for maximal covering location for selecting optimal location of bank branches. So, the results from this research will improve the facility location by GIS and maximal covering location and provide guidelines to get more opportunities to control and monitor Iran's future developments in the locating facilities problem, in the meantime taking into consideration that the city is constantly changing in terms of population, the central district's surrounding competitors, and its accessibility, including the required travel distances and times.

## 3. Methodology

### 3.1. Process

In this article we get criteria from literature review and applying ANP to obtain criterion weight. In the next step, for GIS analysis, the GIS software Esri Arc GIS for Desktop 10 and various GIS layers would be used. The GIS layers represent each of the criteria to be used in the GIS analysis. To manage the analysis, it has to be made sure that sufficient GIS data are available, otherwise develop the GIS layers. While map making preparation, multiple geographic layers are aggregated to produce maps that show the suitability of the land for making the new bank branch. Finally, maximal covering location problem (MCLP) would be applied to select the branches that maximum demand might be reached within a pre-specified target travel time.

### 3.2. Case Study

The study area is Isfahan metropolis, the capital of Isfahan Province, Iran. Bank incident data for a period 1st January 2014 to 31st June 2015 were supplied by the Municipality where each GIS records contain the density of each criterion.

Techniques and approaches of existing standards have not been highlighted any type of aspect required for site selection. This issue highlights the identification of various aspects that may facilitate construction professionals in locating facilities appropriately. It occupies an area of approximately 551 km$^2$. The area extends from 30°43'30" north to 34°27'30"east. Also, Ansar bank is one of the private Iranian banks. We research to suggest an algorithm to locate new bank branch and simulate this model for Ansar bank in Iran metropolis.

The methodology for site selection considers layers and criteria that influence the location of a building. Figure 2 illustrates that the available process contains six steps to identify and assess the results of these aspects. The suitable building locations are general areas where decision makers are satisfied. All processes are conducted to select a general area that is further narrowed down into a few identifiable locations. As already discussed, the selected sites are satisfied. Figure 1 shows the working model of the developed GIS-based methodology.

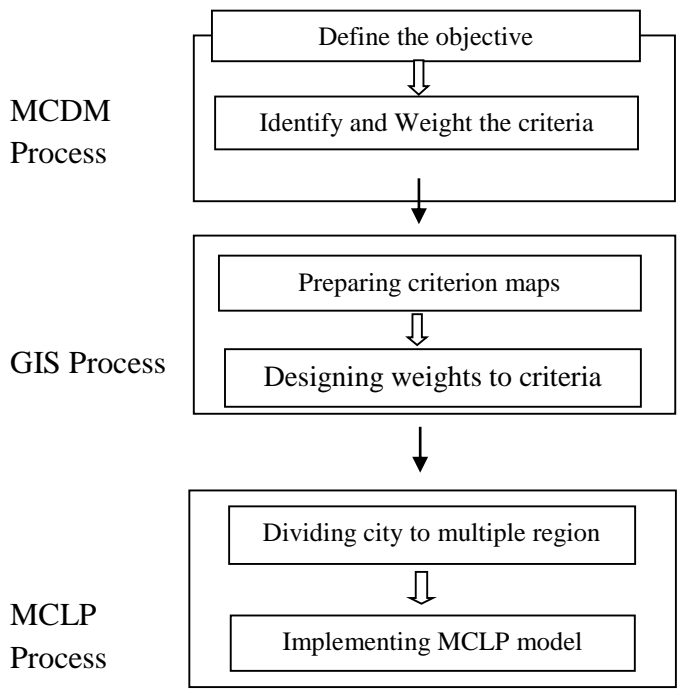

Figure1. The methodology

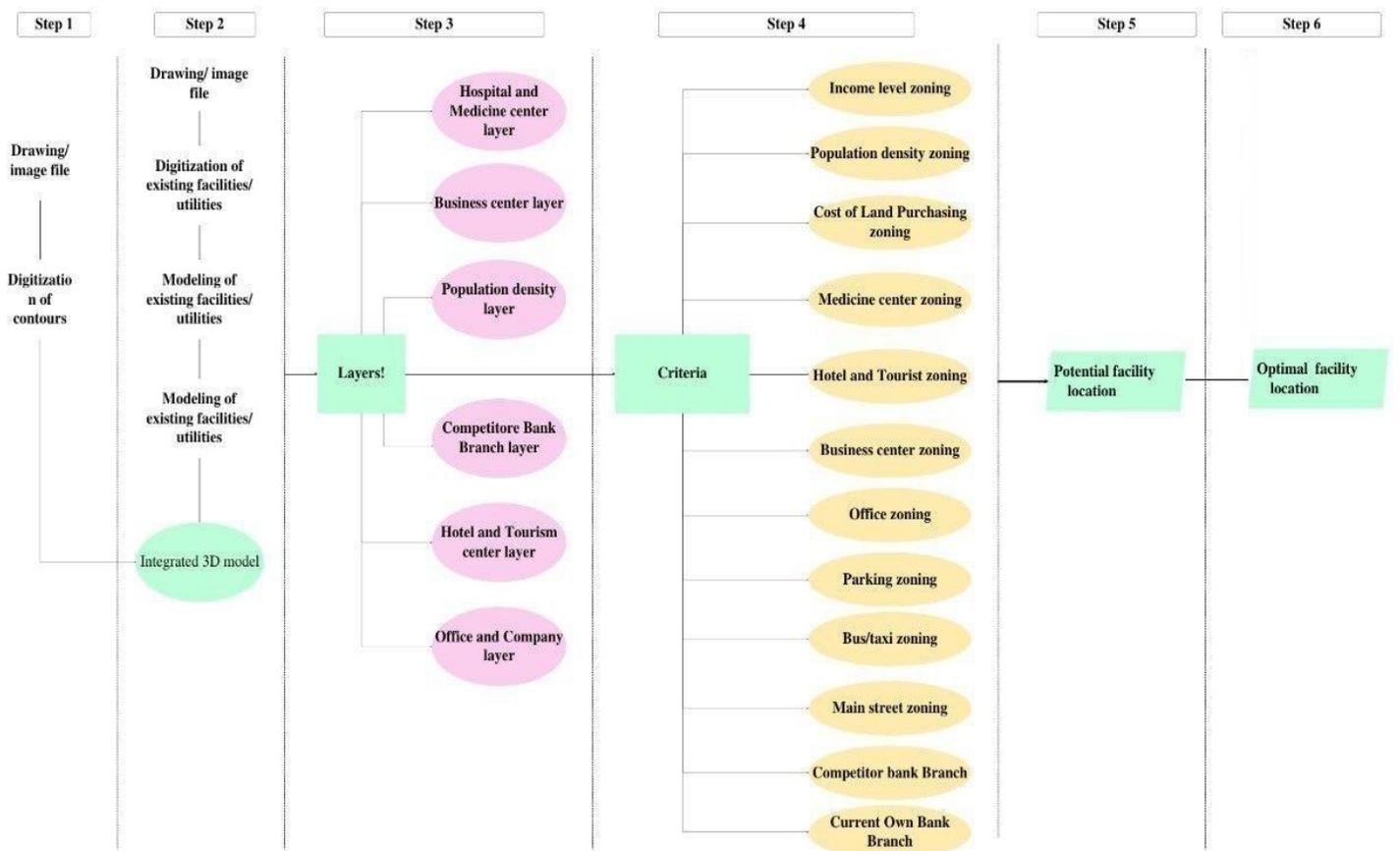

Figure 2. GIS-based model, developed to implement the methodology.

## 4. Define the Objective

The first stage in managing a multi-criteria evaluation analysis is to define the objective of the analysis. In this research, the objective is to present a model for locating the new bank branch.

### 4.1. Identify and Weight the Criteria

In this step Important criteria in locating the new bank branch were determined in six clusters and twelve criteria including: Population Characteristic (Population density, Income level), cost (Cost of Purchase or rent branch's build), urban facilities accessibility (Hospital and medicine vicinity, Business center vicinity, Hotel and tourism center vicinity, Office and company vicinity, Parking vicinity), transportation system (taxi/bus stop and metro/mono rail vicinity, Main Way vicinity), and competition (competitor branch vicinity). Also, weights of each criteria were determined by a group of municipal of staff using the pair wise comparison matrix of the Analytic Network Process. The inconsistency rate of the pair wise comparison matrix was controlled to be less than 0.1.

## 4.2. Preparing Criterion Maps

After collecting Isfahan City GIS map from municipality, the criteria layer would be extracted. Second, the GIS vector layers were made for all of the eleven considered criteria. Third, all of the vector layers would be covered to raster format since the GIS software used to manage the analysis (Esri ArcGIS for Desktop 10) is a raster-based GIS. Some of the GIS layers are shown in Figures 3 to 8.

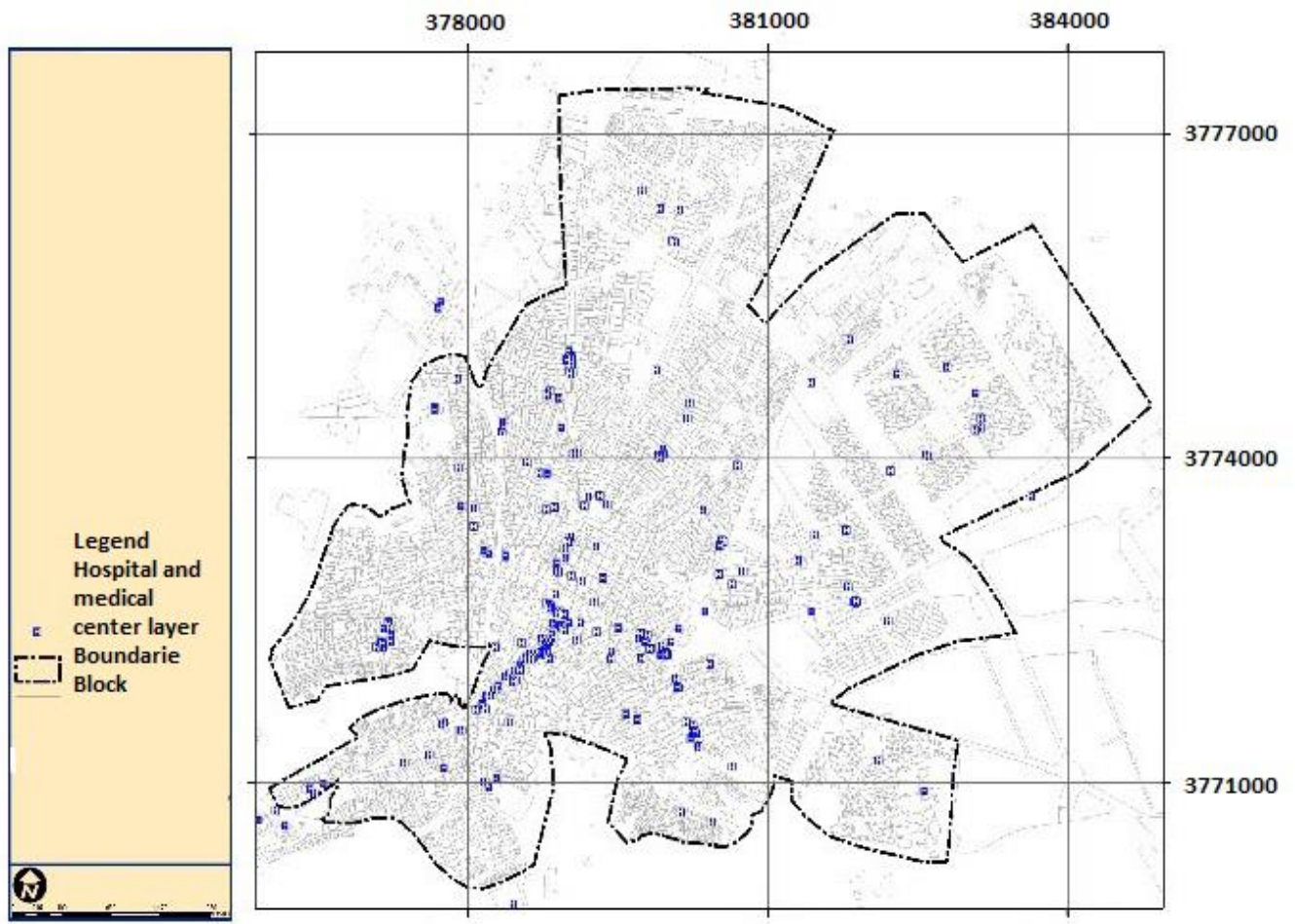

Figure 3. Hospital and Medicine center layer.

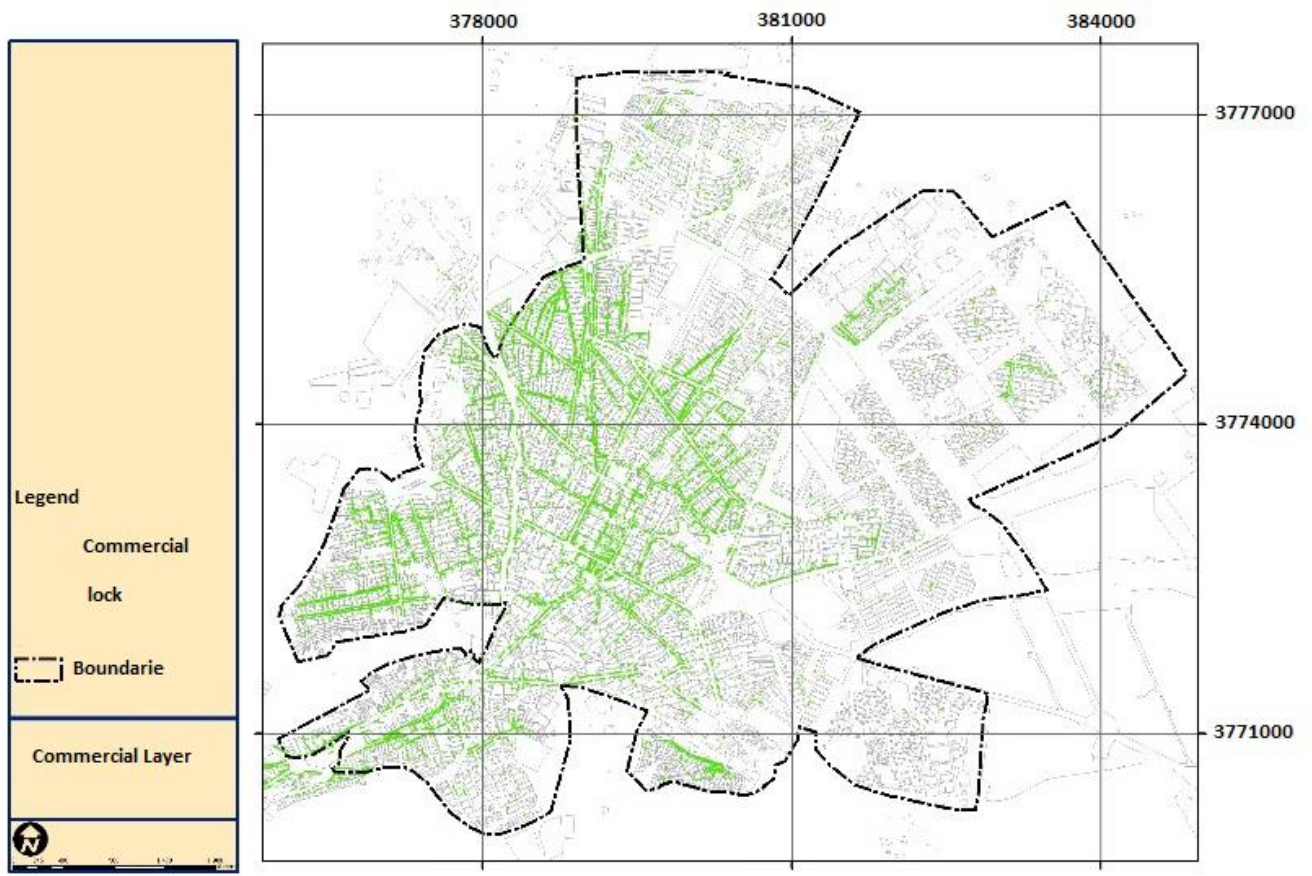

Figure 4. Business center layer.

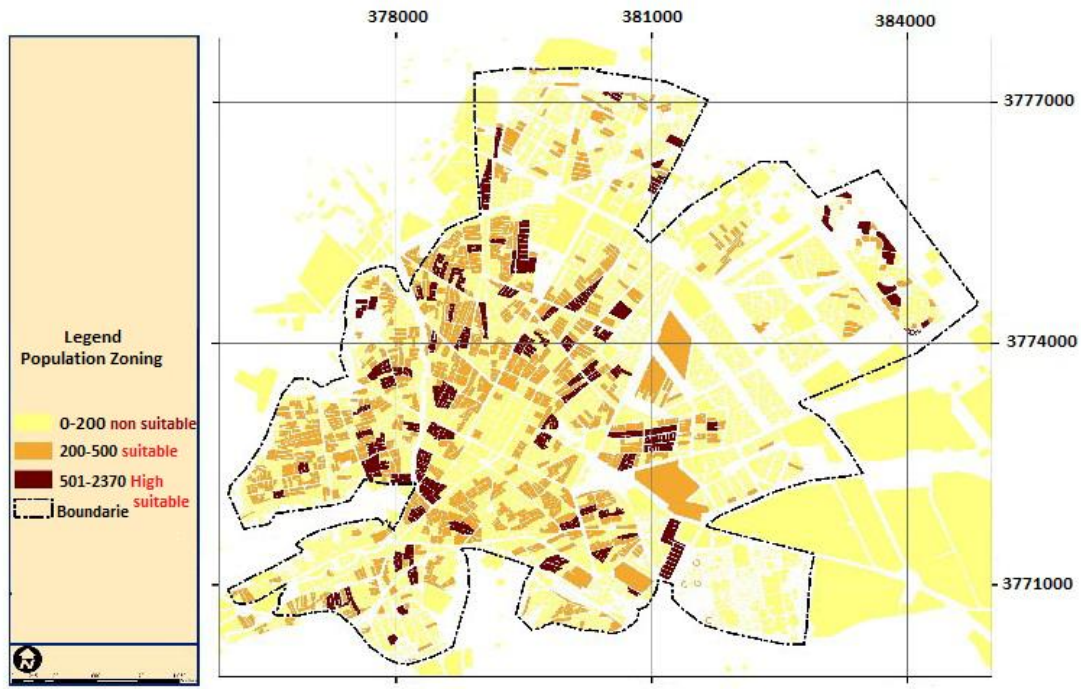

Figure 5. Population density layer.

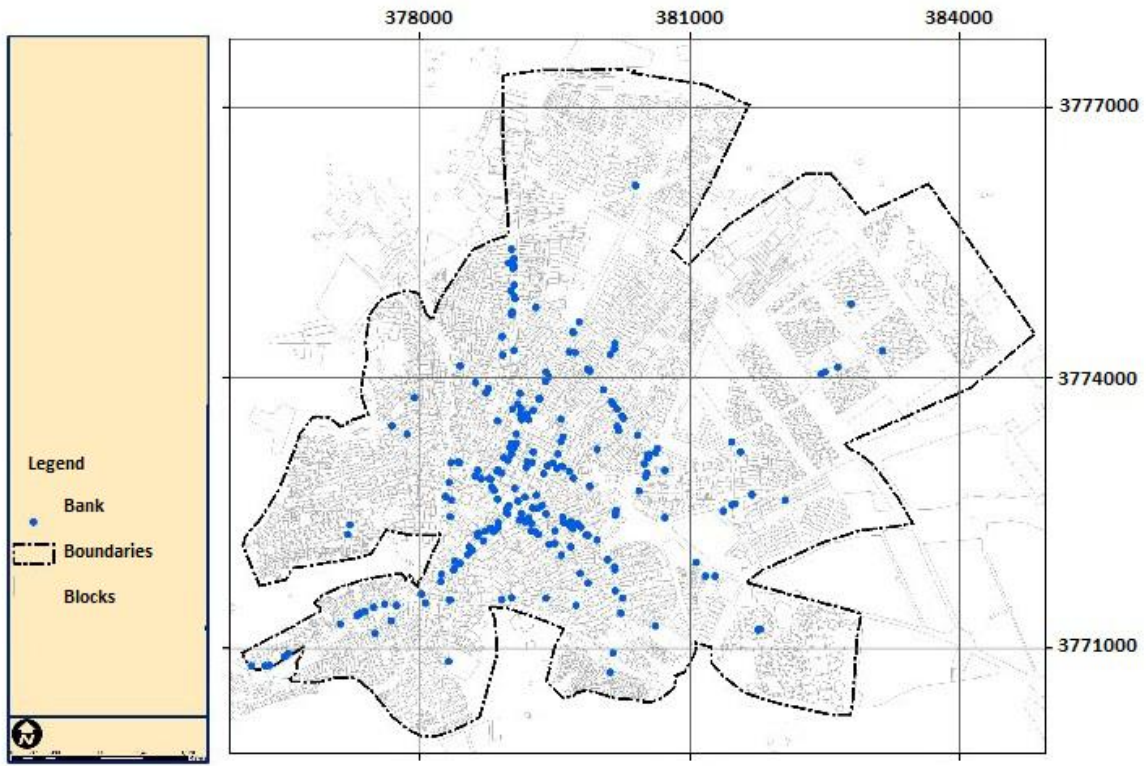

Figure 6. Competitor bank branch layer.

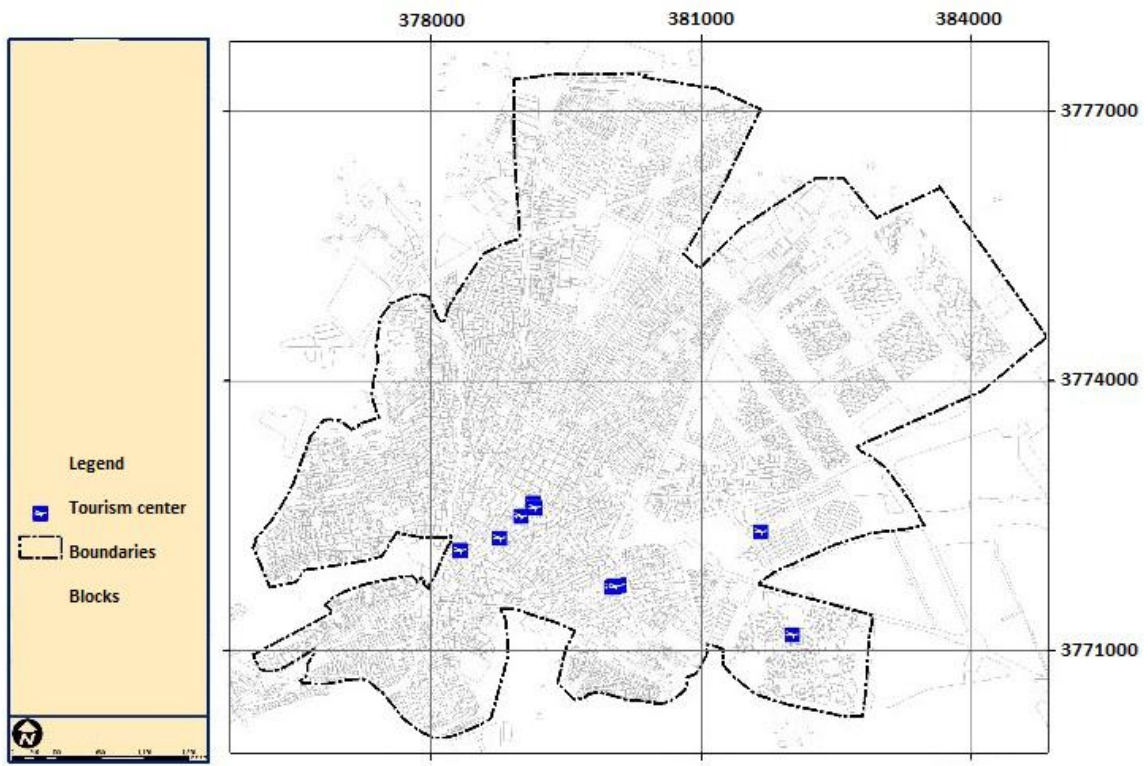

Figure 7. Hotel and tourism center layer.

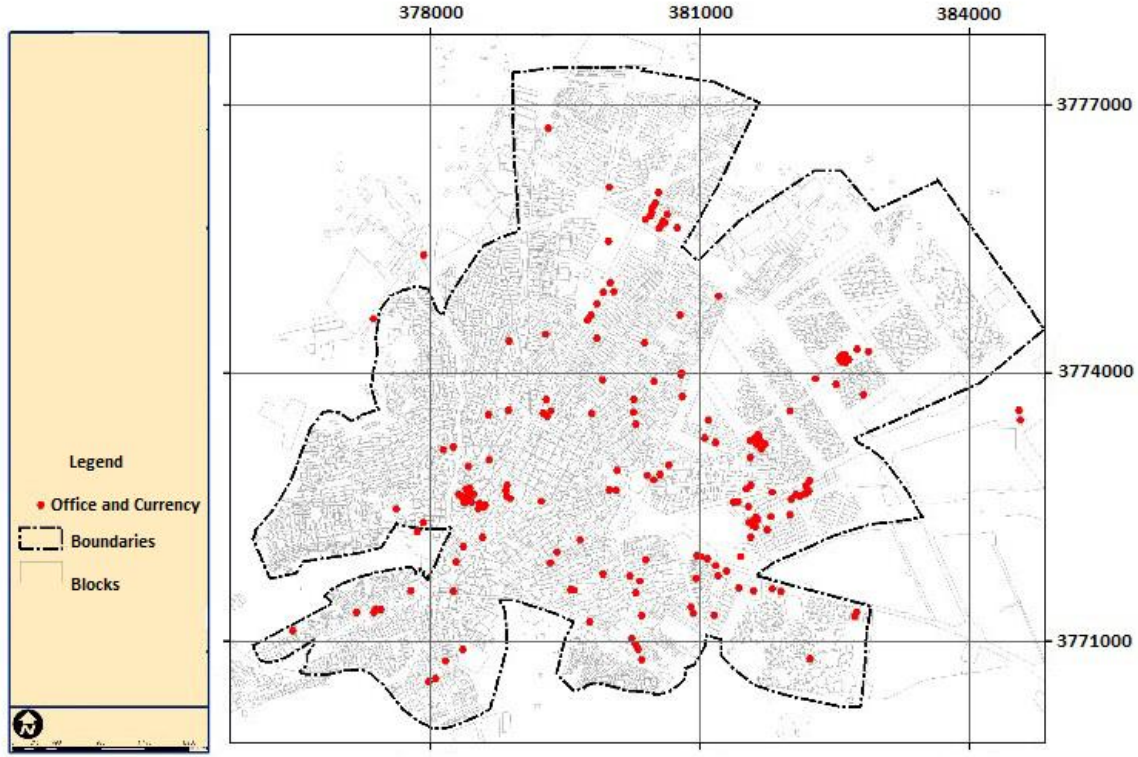

Figure 8. Office and company layer.

## 4.3. Designating Weights to Criteria Maps

Next step after analyzing GIS maps Layer is allocating weights to criteria maps in the GIS software. We used the Esri Arc GIS for Desktop 10 to conduct the analysis and produce the map. The following steps, which are the technical steps, have been done according to Decision Support Wizard of Arc GIS.

The calculated weights were assigned to layers, and different layers of information were combined in ArcGIS software. The hierarchical decision-making structure for location problem is very popular among decision makers. For example, Figure 9 demonstrates the hierarchical decision-making structure consists of 4 layers, which are the main goals, 6 criteria, and 14 sub-criteria and layers.

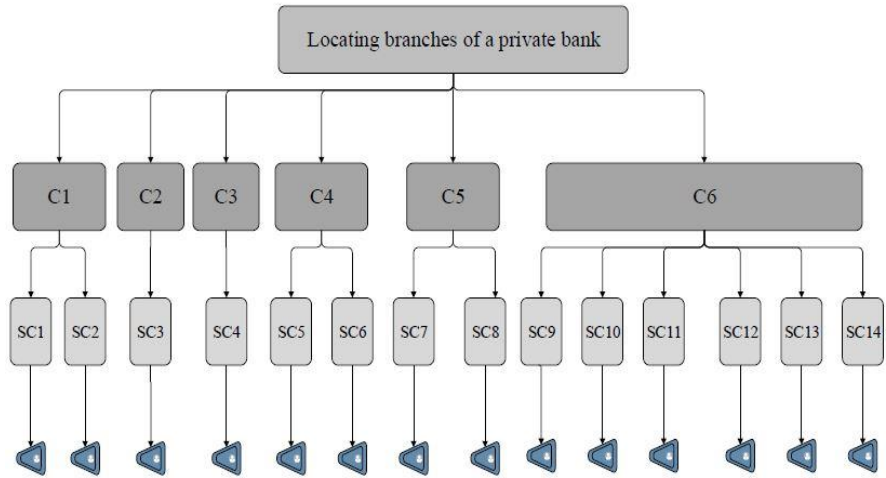

Figure 9. Hierarchical decision-making structure for the location problem.

According to ANP approach, the paper defines the weight of each criteria. To evaluate the located facilities, the paper demonstrates three classes: high suitable class, suitable class, and non-suitable class, and the value of them are 0.6, 0.4 and 0, respectively. According to Table 1, three classification values calculate the location of branches which are defined by using ARCGIS software. Finally, the importance of criterion that obtained in the previous step (Figure 9) is multiplied at its layer, and then all layers are multiplied at each other to getting the best locations for making the new branch. The criteria are shown in Figures 10 to 21.

Table 1. Classification of criterion distance to own bank branch.

| Criteria \ suitability | High suitable | Suitable | Non suitable |
|---|---|---|---|
| Main street | ≤100m | 100-500 m | ≥500 m |
| Business center | ≤100m | 100- 250 m | ≥250 m |
| Hotel and tourism center | ≤1000m | 1000-3000 m | ≥3000 m |
| Office | ≤250m | 200-500 m | ≥500 m |
| Competitor branch | 100-200 m | ≥200m | ≤100 m |
| Familiar branch | ≥1000m | 500-1000m | ≤500 m |
| Income level | High | Middle | Low |
| Cost of purchasing building | Middle | High | Low |
| Medicine center | ≤100m | 100-500m | ≥500 m |

| Population density | ≥500 | 200- 500 | ≤200 |
| Parking | ≤500 m | 500- 1500 m | ≥1500 m |
| Bus/taxi/metro vicinity | ≤500 m | 500- 1500 m | ≥1500 m |

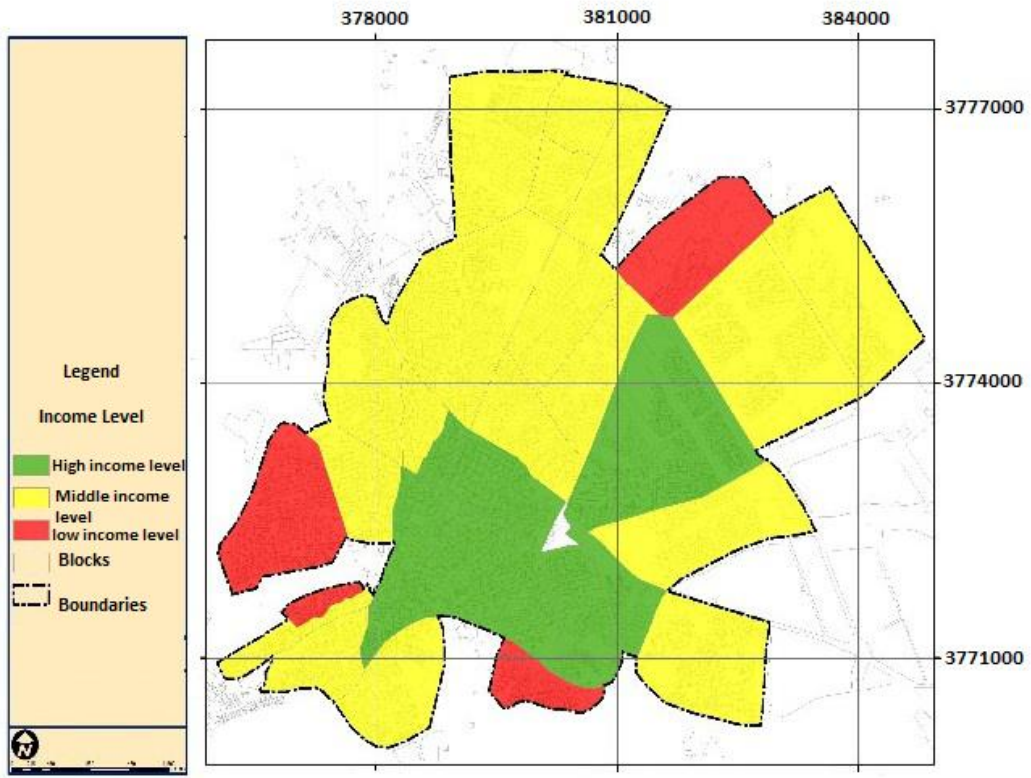

Figure 10. Income level zoning.

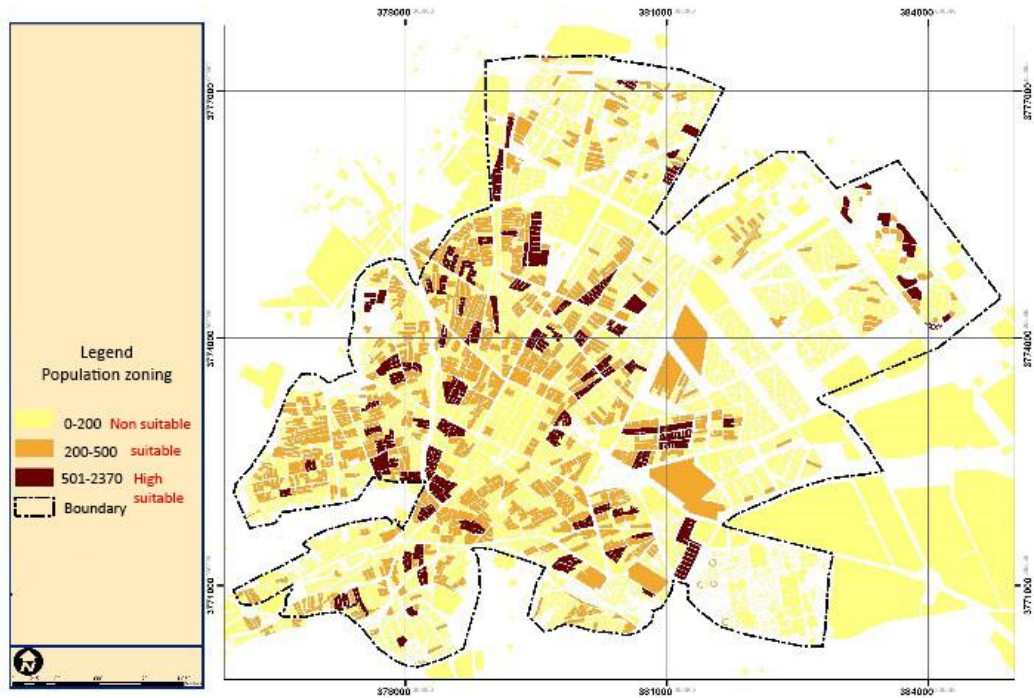

Figure 11. Population density zoning.

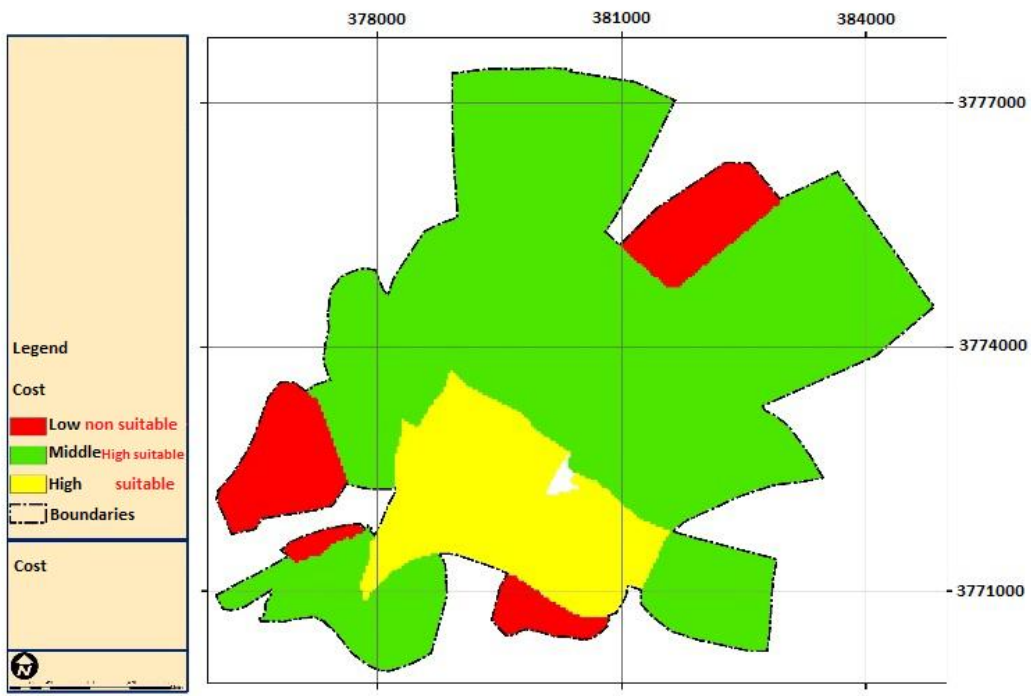

Figure 12. Cost of land purchasing zoning.

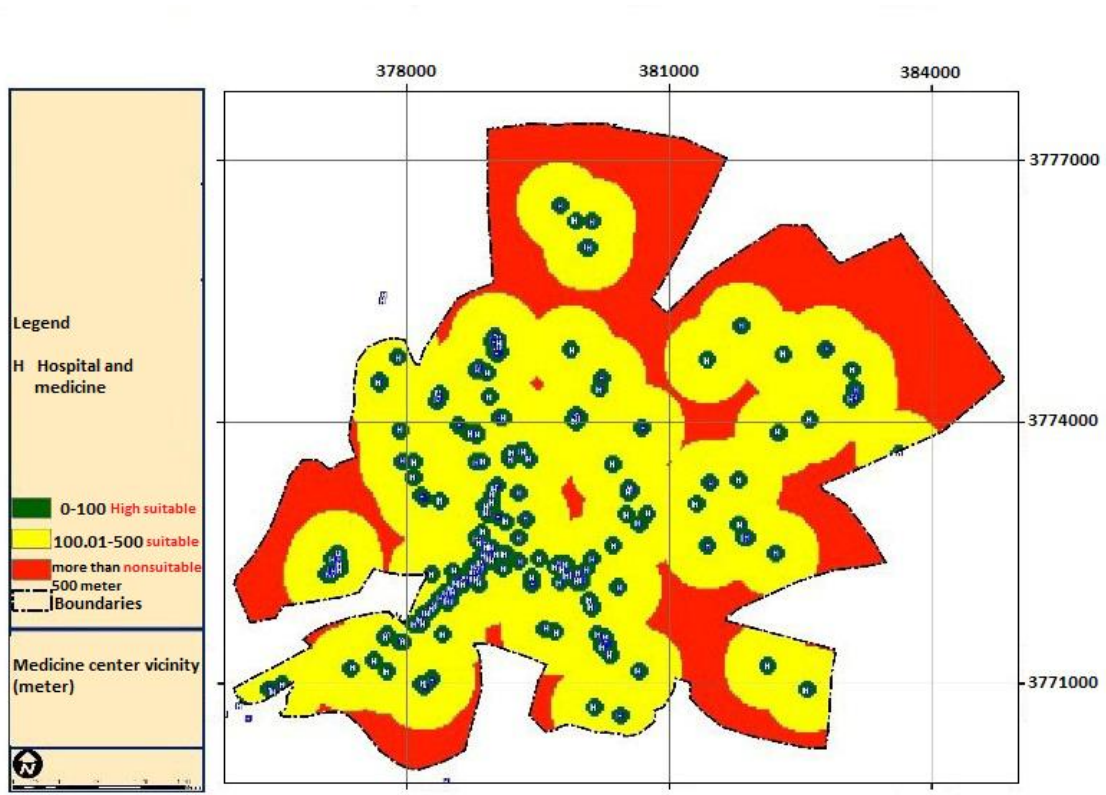

Figure 13. Medicine center zoning.

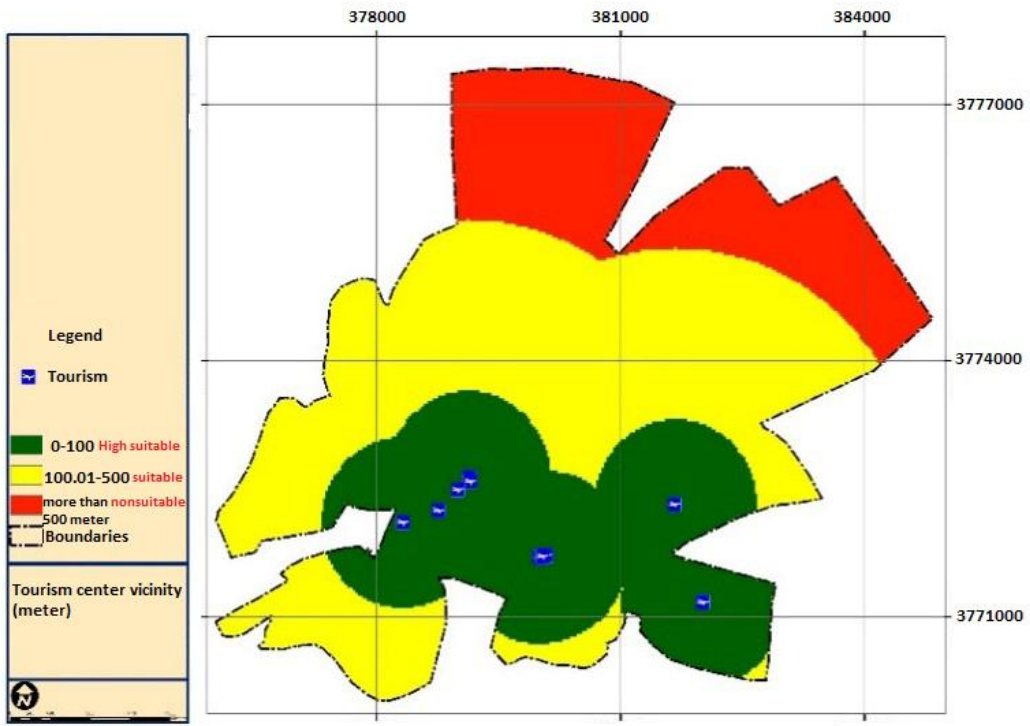

Figure 14. Hotel and tourist.

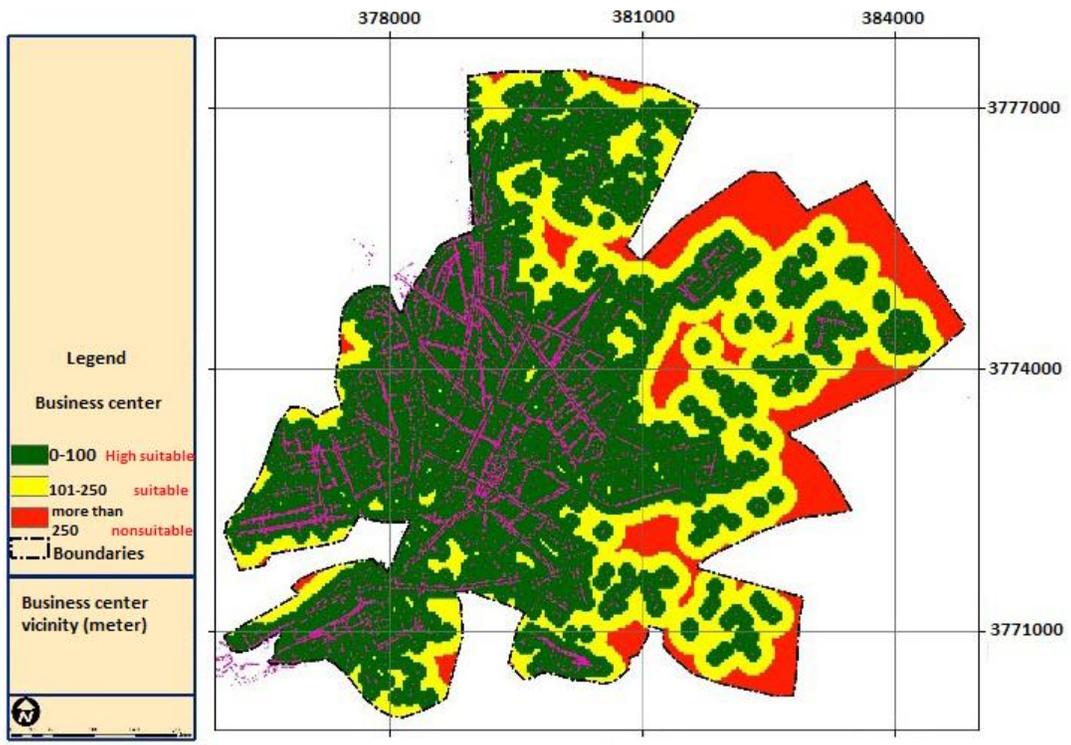

Figure 15. Business center zoning.

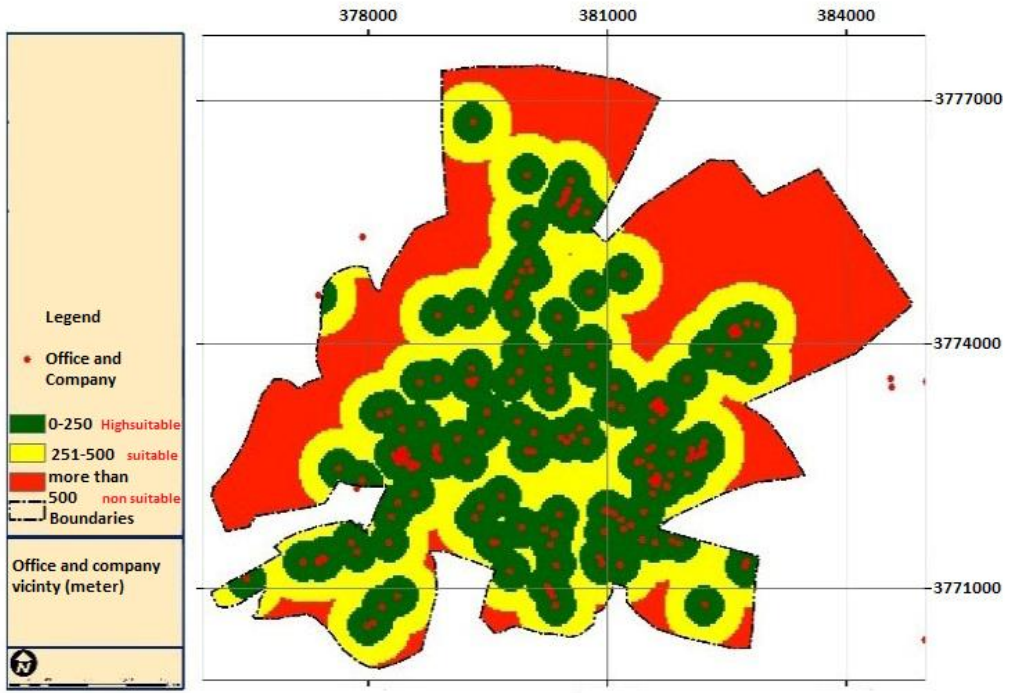

Figure 16. Office zoning.

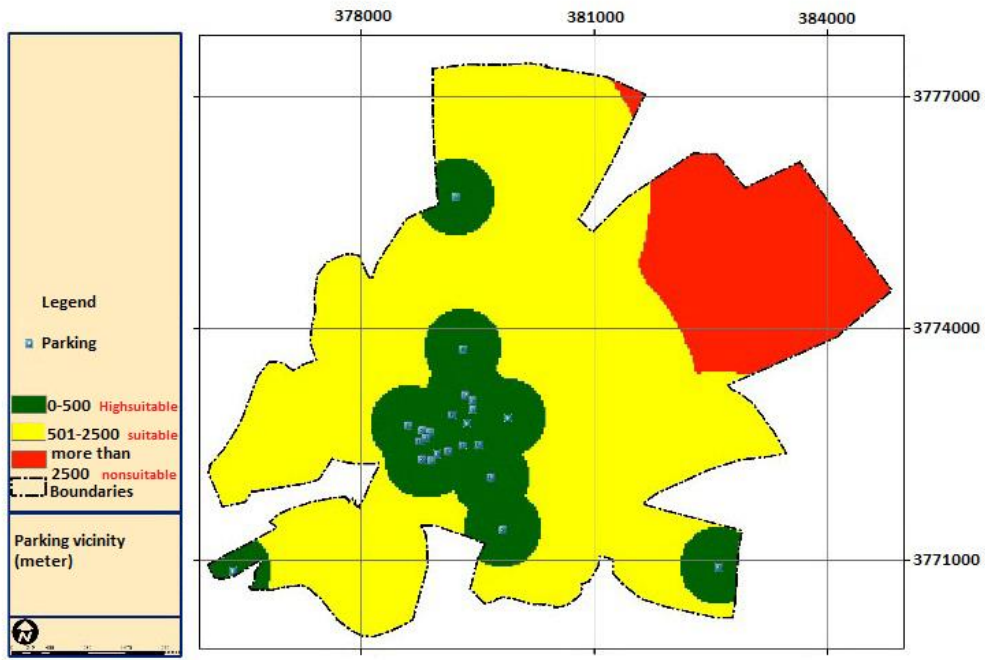

Figure 17. Parking zoning.

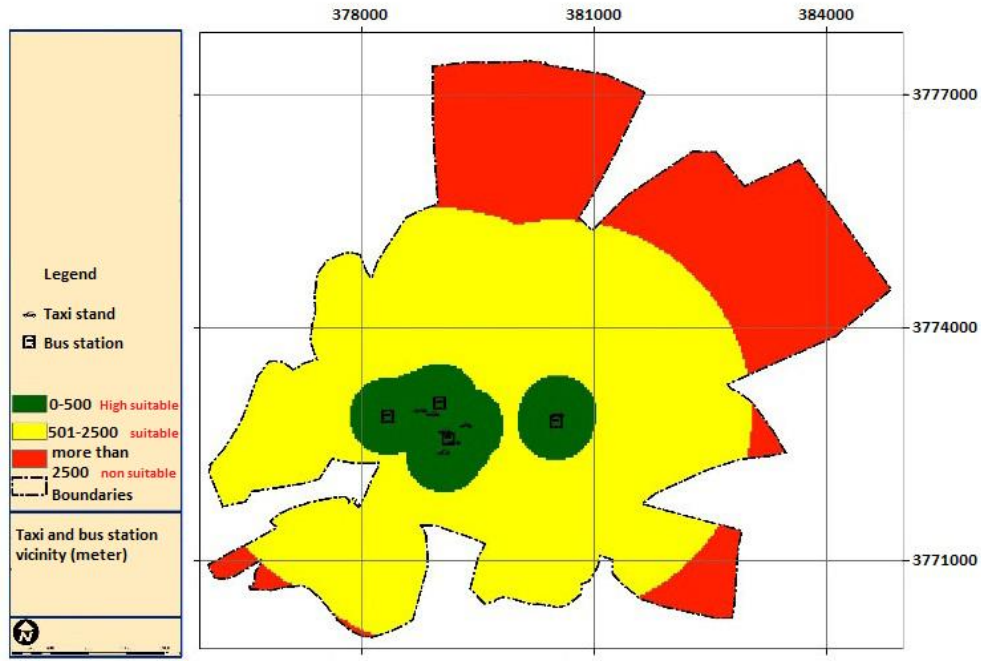

Figure 18. Bus/taxi zoning.

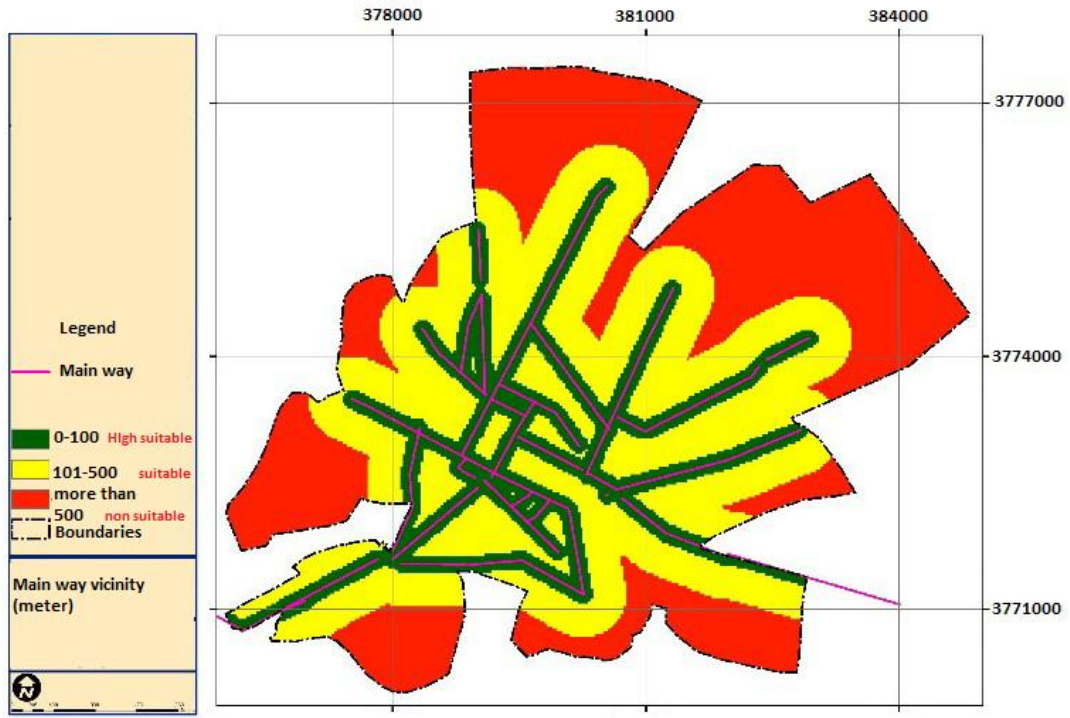

Figure 19. Main street zoning.

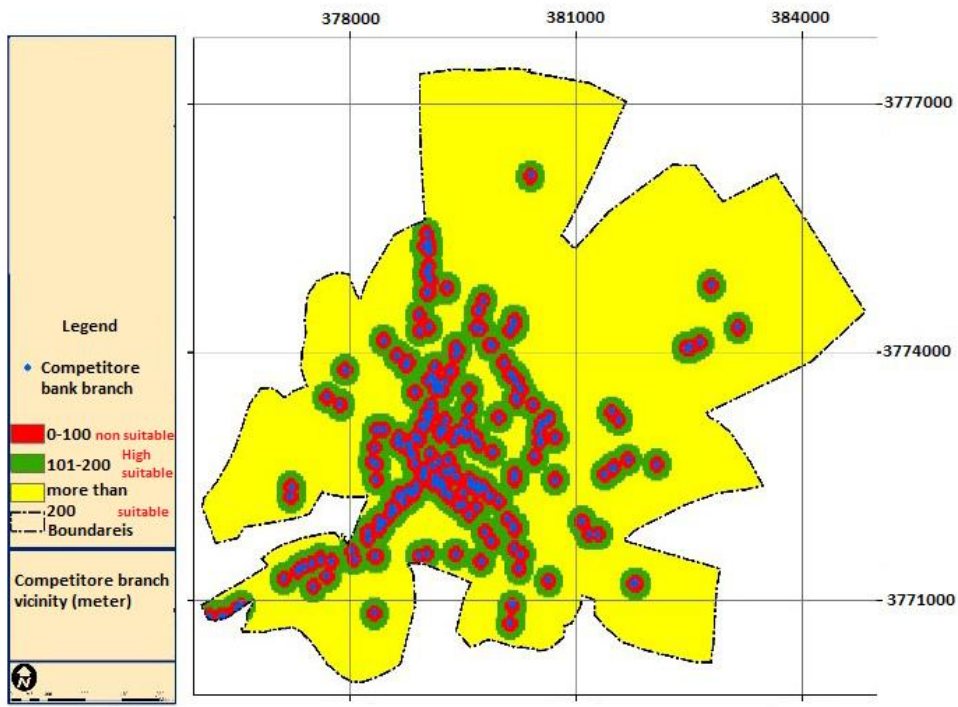

Figure 20. Competitor bank branch.

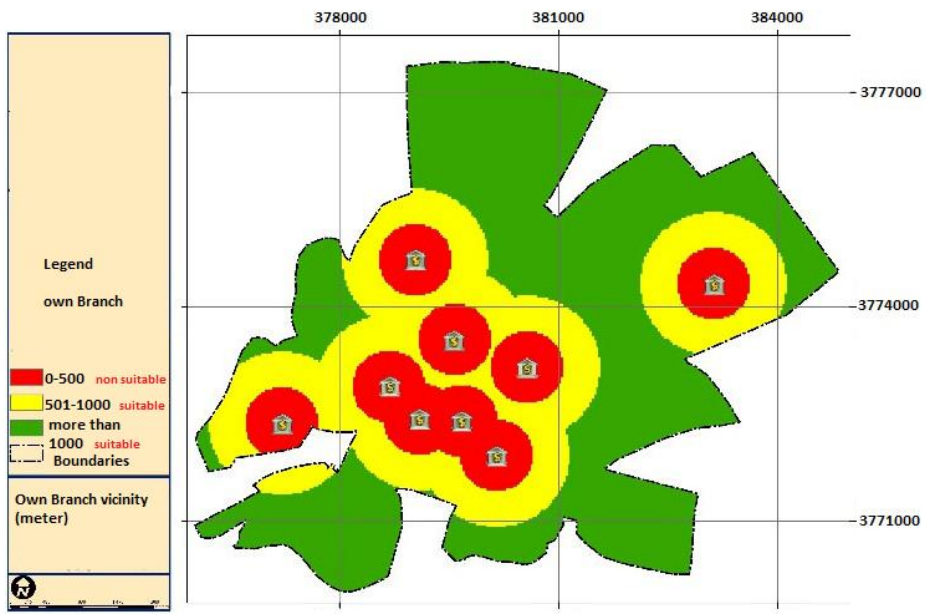

Figure 21. Current own bank branch.

## 4.4. Set of Candidate Sites

After analyzing the decision factors impacting the bank location, twelve criteria in Table 1, contribute to have suitable facility locations. To demonstrate and assess the results, considering Figure 22, the criteria could be depicted. The criteria maps were transferred into point geometry and represented the covering points. All candidate sites were accessible to the state and were literally located. This guarantees that the input (GIS maps) and output (Optimal facility location) can be easily transported. Based on the preliminary selection of the potential sites for bank locations, the potential location of these candidate sites is shown in the map in Figure 23.

After designating weights to criteria map steps, ARC GIS would propose the sites for making the new branch. Also, it classified these sites in three classes including first priority, second priority and third priority according to their scores. There were 23 candidate sites available in the study area, consist of 14 proposed sites and 9 sites of the existing bank branches.

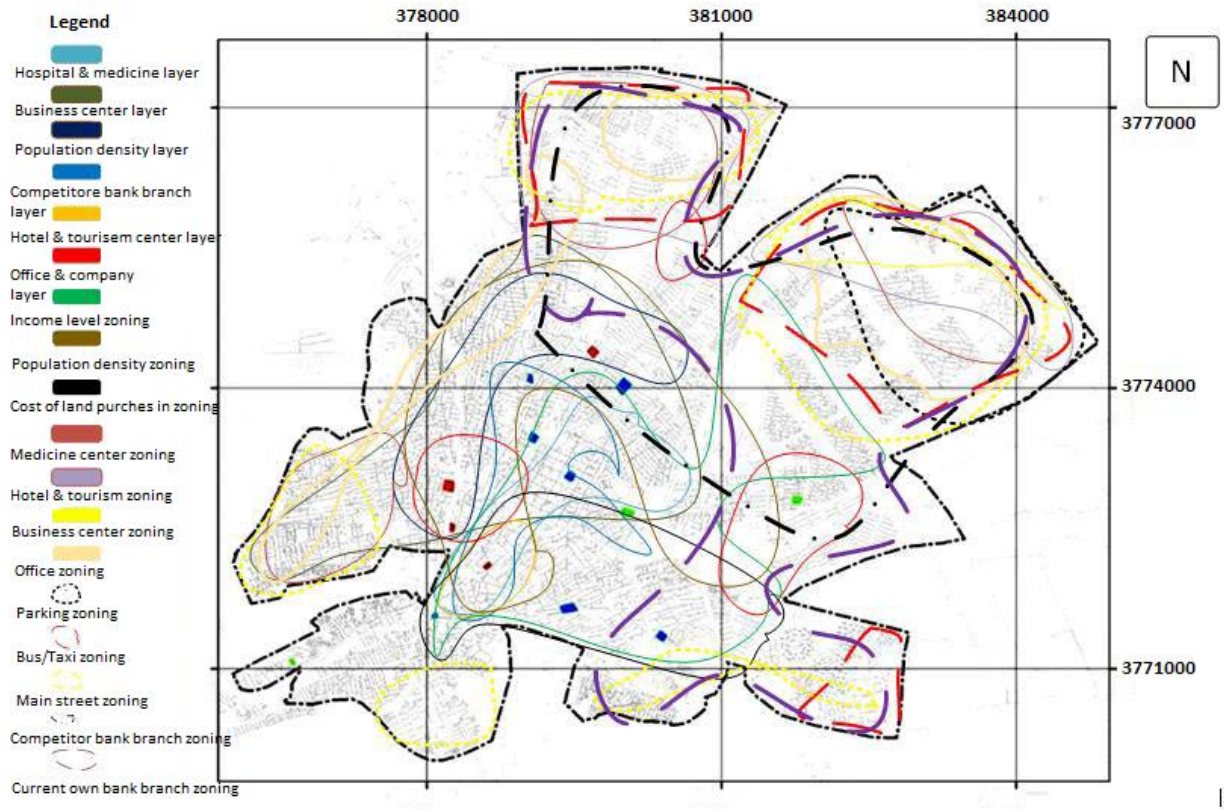

Figure 22. The overlap parts of various factors.

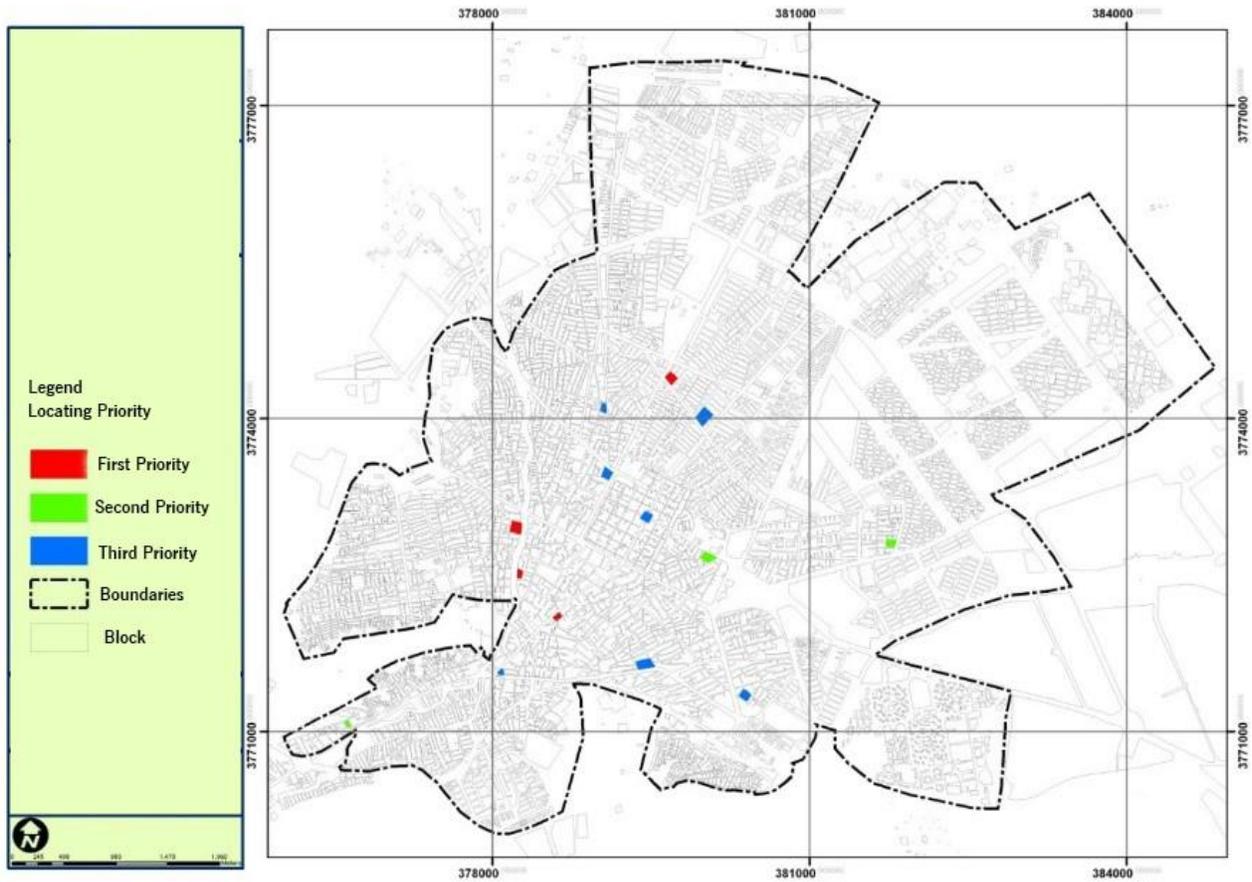

Figure 23. Proposed sites for making the new branch by ARCGIS software.

## 5. Using Maximal Covering Location Problem

This paper also proposes an extension of CMCLP model for locating the bank branches in Isfahan, Isfahan, Iran. This model seeks for maximizing the covered demand (minimizing the uncovered demand). In this problem, each demand area can receive service from several facilities.

In this section, the goal is to apply maximal covering location problem. For using MCLP, according to municipality traffic section, the city was divided to twenty section (twenty demand area). A mathematical formulation of this problem can be stated as follows:

Maximize $z = \sum_{i \epsilon I} a_i y_i$

s.t:  $\sum_{j \epsilon N_i} a_{ij} x_j \geq y_i$   for all i $\epsilon$ I

$$\sum_{j \in J} x_j = p$$

$x_j = (0, 1)$      for all $j \in J$

$y_i = (0, 1)$      for all $i \in I$

Where

I = the set of demand area

J = the set of candidate branch;

$a_{ij}$ = the coverage coefficient

$$a_{ij} = \begin{cases} 1 & \text{if demand area } i \text{ covered by candidate branch } j \\ 0 & \text{otherwise} \end{cases}$$

$$y_i = \begin{cases} 1 & \text{if demand area } i \text{ is covered by at least one candidate branch} \\ 0 & \text{otherwise} \end{cases}$$

$$x_j = \begin{cases} 1 & \text{if a candidate branch alocated to demand area } j \\ 0 & \text{otherwise} \end{cases}$$

$a_i$ = population to be served at demand area i;

p = the number of facilities to be located (number of branches can be located).

After formulating the problem, it was solved by using Lingo software. The results of the problem show the selected branches and covering percentage of the selected locations for branches in Table 2.

Table 2. The optimal solution of MCLP.

| P | Selected location of branches | Covering percentage |
|---|---|---|
| 1 | $x_p$ | 90 |
| 2 | $x_t, x_{pm}$ | 96 |
| 3 | $x_n, x_m, x_t$ | 1 |

As shown in Table 2, approximately 90% of the study area can be covered with making one branch ($x_p$) and nearly 96% with 2 branches ($x_t, x_{pm}$) and 100% of demand area will be covered by 3 branches. Figure 24 exhibits decreasing marginal coverage with each additional new made branch. In other words, the additional coverage obtained by adding the $k_{th}$ branch is generally less than the additional coverage that is obtained by adding the $(k+1)_{th}$ branch.

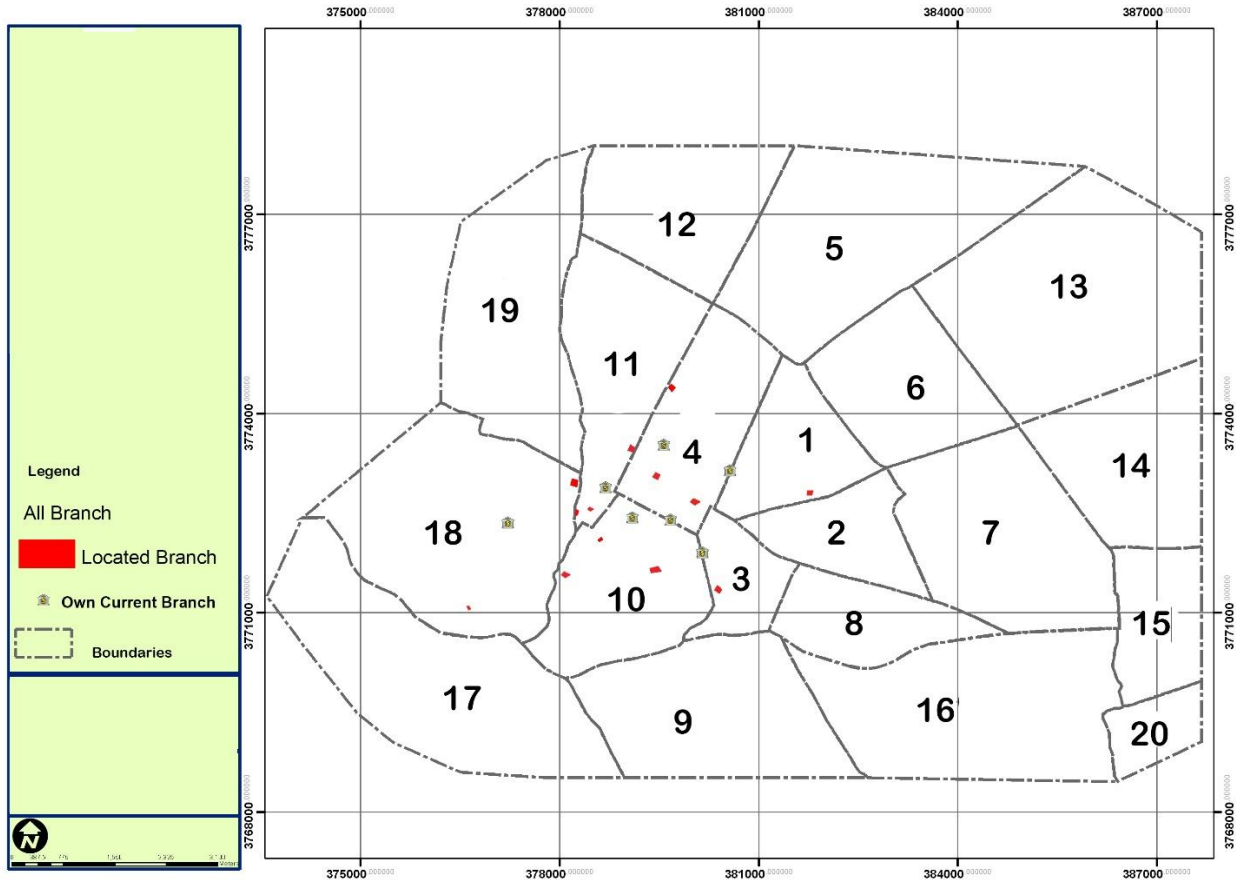

Figure 24. The optimal locations by MCLP.

## 6. Discussion

Since the city of Isfahan suffers from the random allocation of bank branches as well as jammed clustering of the branches, oversupply in some areas of the city has been noticed. The issue is because of the lack of established planning criteria and the absence of careful location selection of the bank branches. This assumption was confirmed after the analysis based on the current study.

Based on the selected location of these two methods, the branch location in both GIS and maximum covering location methods are approximately the same. For analyzing the data GIS was used that detailed datasets allowed relatively accurate and realistic calculations of the significant criteria in the area under the study. Based on the results, the absence of appropriate and feasible planning criteria in the bank branches in Isfahan was confirmed. This problem has led to an oversupply in certain areas, that has resulted in inefficient investments and provision of services in Isfahan. It is clear in the selected area and population ratio that fails to abide with urban planning criteria set in many countries.

## 7. Implications

The results of the current study can be implied for managerial policy making as well as for theoretical perspectives. Based on the results, it was found out that there are excessive numbers of banks in the area under the study at certain locations. Therefore, future bank developments have to be located outside the most covering areas. Other cities and governments around the world might suffer from similar issue. Therefore, they can consider the current study for further site selections for developments of bank branches.

The paper is novel since such studies have never been done in Iran and most countries. Also, this study suggests how to choose the optimal and best location for a branch. The absence of appropriate planning criteria that can regulate the site selection will lead to serious consequences such as oversupply in the bank and negative effects on the performance.

It is vital for educated decision-making to collect accurate, detailed, and updated statistical demographics and socioeconomic information about Isfahan and its neighborhoods. The influence of surrounding developments on the city and its trade situation has significant effects on evaluation of the situation and imperative selected factors. These considerations would definitely help in proper distribution of services. These data have to be reflected on accurate, updated, and detailed GIS maps to provide accurate planning decisions. However, in the bank sector, such decisions can only be achieved after setting various criteria to select sites, according to which strict regulations can be issued that ensure proper coverage of all areas to serve the city's customers and residents. The method used in the paper can be also used for developing other planning criteria; however, it has to be refined to plan all other services such as institutional facilities, transportation systems, social facilities, green areas, health and care services, residential areas, etc. This kind of planning can encourage social investment of the community to obtain a high level of sustainability and ensure the satisfaction of the public with the city's strategic plans.

## 8. Limitations

This study faced with some limitations that the main one was the lack of syndicated data available for the researchers in terms of other banks and the branches of Ansar bank. Consequently, such limitations shifted the aim of the research from business-oriented and mathematical choice modelling towards GIS based methods. In addition, the other limitation of the study was the lack of unified and agreed-upon models for site selection. However, by considering various other factors, including income level zoning, population density zoning, cost of land purchasing zoning, medicine center zoning, hotel and tourist zoning, business center zoning, office zoning, parking zoning, us/taxi zoning, main street zoning, competitor bank branch, and current own bank branch, different results might be obtained.

Furthermore, the lack of the official local and bank services produced potential points of controversy regarding the evaluation criteria against which the situation was refereed. The GIS maps are not available in accepted quality to use in paper to conclude and show clearer. That is, the effect of a bank's tertiary catchment on another's primary or secondary. Correspondingly, the selection site models provide an opportunity that can be filled by future scholars, in an attempt to provide a more realistic.

**Conclusion**

This paper studied the locating bank branches in Isfahan, Iran, based on the defined layers and criterion by using GIS maps. Therefore, the assumption of GIS is one of the most significant decision-making locating processes for banks. Because of the high level of complexity, several factors have to be taken into consideration in the decision-making process while it is the crucial decision. So, this research proposed integration between MCDM model represented by ANP with GIS to propose the best location for construction of the new bank branch, and then using Maximal Covering Location Problem to select the branches that the maximum demand might be reached within a pre-specified target travel time. The model was implemented for Ansar bank in Isfahan city, Iran. As a result of the research, it was found out that the proposed model is practical and effective for identifying suitable sites with respect to multiple criteria. Therefore, the proposed model could be used for locating the popular and accessible sites in cities such as locating health care centers, emergencies and hospitals, suppliers and factories, police stations and taxi and bus stations. The model used a combination approach of operation research and GIS instead of proposing some location empirically and ranking them with MCDM techniques. For future researches, it is recommended to use study reconstruction of bank branch as well as applying this model on a more complex problem such as measuring the merits of locating.

**References**

Adil, M. (2013) 'Modelling effect of perceived service quality dimensions on customer satisfaction in Indian bank settings.' *International Journal of Services and Operations Management*, Vol. *15* No 3, pp. 358-373.

Alsalloum, O.I., and Rand G.K. (2006) 'Extensions to emergency vehicle location models. *Computers and Operations Research*, 33(9), pp. 2725 – 2743.

Amukele, T., Ness, P. M., Tobian, A. A., Boyd, J., & Street, J. (2017) "Drone transportation of blood products." *Transfusion*, Vol. 57 No.3, pp. 582-588.

Atta, S., Mahapatra, P. R. S., and Mukhopadhyay, A. (2018) 'Solving maximal covering location problem using genetic algorithm with local refinement.' *Soft Computing*, Vol. *22* No. 12, pp. 3891-3906.

Bansal, M. (2018). On Solving General Planar Maximum Coverage Location Problem with Partial Coverage, Optimization-Online. *https://pubsonline.informs.org/doi/abs/10.1287/ijoc.2016.0722*, (3 January 2017).

Bansal, M., and Kianfar, K. (2017) 'Planar maximum coverage location problem with partial coverage and rectangular demand and service zones.' *INFORMS Journal on Computing,* Vol. 29 No.1, pp. 152-169.

Berman, O., Drezner, Z., and Krass, D. (2019) 'The multiple gradual cover location problem.' *Journal of the Operational Research Society*, Vol. 70. No. 6, pp. 931-940.

Boutilier, J. J., Brooks, S. C., Janmohamed, A., Byers, A., Buick, J. E., Zhan, C., and Chan, T. C. (2017) 'Optimizing a drone network to deliver automated external defibrillators.' *Circulation*, Vol. *135* No. 25, pp.2454-2465.

Chaudhary, S., Nidhi, C., and Rawal, N. R. (2019) 'GIS-Based Model for Optimal Collection and Transportation System for Solid Waste in Allahabad City.' In *Emerging Technologies in Data Mining and Information Security* (pp. 45-65). Springer, Singapore.

Chauhan, D., Unnikrishnan, A., and Figliozzi, M. (2019) 'Maximum coverage capacitated facility location problem with range constrained drones.' *Transportation Research Part C: Emerging Technologies*, Vol. *99*, pp. 1-18.

Cheng, E. W.L., Li H., and Yu L. (2007) "A GIS approach to shopping mall location selection." *Building and Environment*, Vol. 42, pp. 884–892.

Claesson, A., Bäckman, A., Ringh, M., Svensson, L., Nordberg, P., Djärv, T., & Hollenberg, J. (2017) "Time to delivery of an automated external defibrillator using a drone for simulated out-of-hospital cardiac arrests vs emergency medical services." *Jama*, Vol. 317 No. 22, pp. 2332-2334.

Cordeau, J. F., Furini, F., & Ljubić, I. (2019) "Benders decomposition for very large scale partial set covering and maximal covering location problems." *European Journal of Operational Research*, Vol. 275 No. 3, pp. 882-896.

Dick, A. A., 2008 'Demand Estimation and Consumer Welfare in the Banking Industry,' *Journal of Banking and Finance,* Vol. 32 No. 8, pp. 1661–1676.

Ernst, A.T., Jiang, H., Krishanmoorthy, M., and Baatar, D. (2018) 'Reformulations and Computational Results for the Incapacitated Single Allocation Hub Covering Problem. In: Sarker R., Abbass H., Dunstall S., Kilby P., Davis R., Young L. (eds) Data and Decision Sciences in Action. Lecture Notes in Management and Industrial Engineering. Springer, Cham, pp. 133-148.

Fung, A. (2001). Half of Consumers Still Prefer Branches, Poll Finds. *American Banker*, 17.

García Cabello, J., (2017) 'A decision model for bank branch site selection: Define branch success and do not deviate,' *Socio-Economic Planning Sciences*, pp. 1-10.

Garegnani, G., Sacchelli S., Balest, J., and Zambelli, P., (2018) 'GIS-based approach for assessing the energy potential and the financial feasibility of run-off-river hydro-power in Alpine valleys.' *Applied Energy*, Vol. 216, pp. 709-723.

Grimshaw, D.J. (1994). Bringing Geographical Information Systems into Business. Harlow: Longman.

Ibrahim, E. H., Mohamed, S. E., and Atwan, A. A. (2011) 'Combining fuzzy analytic hierarchy process and GIS to select the best location for a wastewater lift station in El-Mahalla El-Kubra, North Egypt.' *International Journal of Engineering & Technology*, Vol. 11 No. 5, pp.44-50.

Kamble, S.S., Dhume, S.M., Raut, R.D. and Chaudhuri, R. (2011) 'Measurement of service quality in banks: a comparative study between public and private banks in India', *Int. J. Services and Operations Management*, Vol. 10, No. 3, pp.274–293.

Liu, S., and Hodgson, M. E. (2013) 'Optimizing large area coverage from multiple satellite-sensors.' *GIS cience and remote sensing*, Vol. 50 No. 6, pp. 652-666.

Lyu, H. M., Sun, W. J., Shen, S. L., and Arulrajah, A. (2018) 'Flood risk assessment in metro systems of mega-cities using a GIS-based modeling approach.' *Science of the Total Environment*, Vol. 626, pp. 1012-1025.

Lotfalipour, Z., NajiAzimi, Z., and Kazemi, M. (2014) 'Locating the bank branches using a hybrid method Technical,' *Journal of Engineering and Applied Sciences,* Vol. 4 No. 3, pp. 124-134.

Madani, S. R., Nookabadi, A. S., and Hejazi, S. R. (2018) 'A bi-objective, reliable single allocation p-hub maximal covering location problem: Mathematical formulation and solution approach.' *Journal of Air Transport Management*, Vol. *68*, pp.118-136.

Meyer, E. (2011). Performing Location Allocation Measures with a GIS for Fire Stations in Toledo, Ohio. Master of art degree thesis, The University of Toledo.

Murad, A.A. (2011) 'Using GIS for evaluating retail centres location at Jeddah City', *International Journal of Services and Operations Management*, Vol. 10, No. 3, pp.255–273.

Murray, A. T., Xu, J., Wang, Z., and Church, R. L. (2019) 'Commercial GIS location analytics: capabilities and performance.' *International Journal of Geographical Information Science*, Vol. 33 No. 5, 1106-1130.

Palomino Cuya DG, Brandimarte L, Popescu I, Alterach J, and Peviani M. (2013) 'A GIS-based assessment of maximum potential hydropower production in La Plata basin under global changes.' *Renewable Energy*; Vol. 50 pp. 103–14.

Potter, R.B., Darmame, K., Barham, N. and Nortcliff, S., (2007) An Introduction to the Urban Geography of Amman, Jordan. Geographic Paper No. 182. Geographic Papers, The University of Reading, pp. 1–29

Rajendran, S., (2006) Sustainable Construction Safety and Health Rating System. Oregon State University, Corvallis, Ore.

Rajendran, S., and Gambatese, J.A., (2009) 'Development and initial validation of sustainable construction safety and health rating system.' *Journal of Construction Engineering and Management.* Vol. 135 No. 10, pp. 1067–1075.

Rezazadeh, H., Moghtased-Azar S., and Shafiei Kisomi. S., and Bagheri, R., (2018) 'Robust cooperative maximal covering location problem: a case study of the locating Tele-Taxi stations in Tabriz, Iran,' *International Journal Services and Operations Management*, Vol. 29 No. 2, pp. 163–183.

Shi, L., Meyer, R. R. Bozbay, M. and Miller, A. J. (2004) 'A nested partitions framework for solving large-scale multi commodity facility location problems,' *Journal of System and Science System Engineering*, Vol. 13, No. 2, pp. 158–179.

Tavakoli, M. M., Shirouyehzad, H., and Dabestani, R. (2013) 'Efficiency evaluation of a private bank's branches with service quality approach by data envelopment analysis.' *International Journal of Services and Operations Management*, Vol. 16 No.4, pp. 427-442.

Vahidnia, M. H., Alesheikh, A., Alimohammadi, A., and Bassiri, A. (2008). 'Fuzzy analytical hierarchy process in GIS application.' *The International Archives of the Photogrammetry, Remote Sensing and Spatial Information Sciences*, Vol. 37(B2), pp. 593-596.

Vlachopoulou, M., Silleos, G., and Manthou, V. (2001). "Geographic information systems in warehouse siteselection decisions." *International Journal of Production Economics*, Vol.71, pp. 205-212.

Wang, X.Z., Zhou J., and Huang Z.L., (2016) "A multilevel deep learning method for big data analysis and emergency management of power system." *IEEE international conference on big data analysis. IEEE Press*, pp. 1–5.

Yao, J., Zhang, X., and Murray, A. T. (2019) 'Location optimization of urban fire stations: Access and service coverage.' *Computers, Environment and Urban Systems*, Vol. 73, pp. 184-190.

Zabihi, M., Pourghasemi, H. R., Motevalli, A., and Zakeri, M. A. (2019) Gully erosion modeling using GIS-based data mining techniques in Northern Iran: a comparison between boosted regression tree and multivariate adaptive regression spline. In *Natural Hazards GIS-Based Spatial Modeling Using Data Mining Techniques* (pp. 1-26). *Springer*, Cham.

Zanjirani Farahani, R., Asgari, N., Heidari, N., Hosseininia, M., and Goh, M., (2012) 'Covering problems in facility location: A review.' *Computers and Industrial Engineering*, Vol. 62 No.1, pp. 368-407.